\documentclass[twocolumn,showpacs,showkeys,reprint,amsmath,amssymb,aps]{revtex4-2}

\usepackage{graphicx}
\usepackage{dcolumn}
\usepackage{bm}
\usepackage{tabularx}
\usepackage{amsmath}
\usepackage{soul, color, xcolor}

\soulregister{\cite}7
\soulregister{\ref}7
\soulregister{\pageref}7
\soulregister{\eqref}7

\emergencystretch=\maxdimen
\begin{document}

\title{Pressure-induced structural and superconducting transitions in black arsenic}

\author{Y. Y. Wu,$^{1,2,*}$ L. Mu,$^{1,*}$ Y. L. Zhang,$^{3,4,*}$ D. Z. Dai,$^{1}$ K. Meng,$^{1}$ S. Y. Huang,$^{5,\dag}$ X. Zhang,$^{1}$ S. C. Huang,$^{1}$ J. Chen,$^{4,6,7}$ H. G. Yan,$^{1,8}$ and S. Y. Li$^{1,2,9,10,\ddag}$}

\affiliation{
$^{1}$ State Key Laboratory of Surface Physics, Department of Physics, Fudan University, Shanghai 200438, China\\
$^{2}$ Shanghai Research Center for Quantum Sciences, Shanghai 201315, China\\
$^{3}$ Ningbo Key Laboratory of All-Solid-State Battery, Eastern Institute for Advanced Study, Eastern Institute of Technology, Ningbo, Zhejiang 315200, China\\
$^{4}$ School of Physics, Peking University, Beijing 100871, China\\
$^{5}$ Institute of Optoelectronics, Fudan University, Shanghai 200438, China\\
$^{6}$ Interdisciplinary Institute of Light-Element Quantum Materials and Research Center for Light-Element Advanced Materials, Peking University, Beijing 100871, China\\
$^{7}$ Frontiers Science Center for Nano-Optoelectronics, Peking University, Beijing 100871, China\\
$^{8}$ Key Laboratory of Micro and Nano-Photonic Structures (Ministry of Education), Fudan University, Shanghai 200438, China\\
$^{9}$ Shanghai Branch, Hefei National Laboratory, Shanghai 201315, China\\
$^{10}$ Collaborative Innovation Center of Advanced Microstructures, Nanjing 210093, China
}

\date{\today}

\begin{abstract}

We report high-pressure Raman spectra and resistance measurements of black arsenic (b-As) up to 58 GPa, along with phonon density of states (DOS) and enthalpy calculations for four reported arsenic phases up to 50 GPa. It is found that metastable b-As transforms into gray arsenic (g-As) phase at a critical pressure of 1.51 GPa, followed by subsequent transitions to simple cubic arsenic (c-As) and incommensurate host-guest arsenic (hg-As) phases at 25.9 and 44.8 GPa, respectively. Superconductivity emerges above 25 GPa in the c-As phase, with the superconducting transition temperature ($T${$\rm_c$}) remaining nearly a constant of 3 K. Upon further compression, $T${$\rm_c$} steeply increases to a higher value around 4.5 K in the incommensurate hg-As phase above 43 GPa. We use our results to update the structural and superconducting phase diagrams under pressure for the novel semiconductor, black arsenic.
\end{abstract}

\maketitle

\section{introduction}

After the successful fabrication of few-layer black phosphorus devices \cite{LLK}, the group VA materials (phosphorus, arsenic, antimony, and bismuth) have attracted great attention due to their unique layered structures and electronic properties. At ambient pressure, the group VA materials are either semiconductors or semimetals \cite{ZhangSL, BatoolS}. The application of high pressure serves as a clean and effective method to modify the crystal structure and electronic properties, providing opportunities to explore the novel phase transitions in various materials \cite{ZhangL, FLJA2}. Under pressure, all the group VA materials undergo structural and superconducting transitions \cite{KaruzawaM, GuoJ, FLJA, KawamuraH, ChenAL, BiJW, Ilina}. For example, black phosphorus manifests superconductivity at the critical pressure 5 GPa of the orthorhombic-rhombohedral structural transition, with a superconducting transition temperature ($T${$\rm_c$}) of 3.2 K \cite{GuoJ}. As pressure further increases, $T${$\rm_c$} gradually increases and finally saturates at $\sim$10 K above 30 GPa \cite{KaruzawaM, GuoJ, FLJA}.

Metastable black arsenic (b-As) is isomorphic to black phosphorus, can also be exfoliated to few layers, and exhibits interesting electronic properties \cite{ZhongM, ZhongM2, ChenY, WangY, ShengF}. The bandgap of bulk b-As is $\sim$0.29 eV with $n$-type doping \cite{STMBAs}. Similar to black phosphorus, b-As has a layer-dependent band gap and remarkable in-plane anisotropic features \cite{ZhongM, ZhongM2, ChenY, WangY}. Few-layer b-As devices demonstrate high carrier mobility, large on/off ratios, and superior air stability compared to black phosphorus \cite{ZhongM, ZhongM2, ChenY}. Furthermore, the spin-orbital coupling and Stark effect in few-layer b-As lead to the formation of particle-hole asymmetric Rashba valley, and manifest exotic quantum Hall states that can be reversibly controlled by electrostatic gating \cite{ShengF}. In this regard, b-As holds great promise as a candidate material for further electronic device applications.

The studies on b-As under pressure have shown that it undergoes a structural transition to gray arsenic (g-As) phase at 3.48 GPa, with both b-As and g-As coexisting between 3.48 and 4.62 GPa \cite{LiRP, Gao2020}. At pressures above 5.37 GPa, b-As completely transforms to g-As phase \cite{Gao2020}, and then to simple cubic arsenic (c-As) phase at about 25 GPa \cite{Du2023}. On decompression, the sample remains in g-As phase after the pressure is released to ambient pressure \cite{Gao2020, Du2023}. Starting from g-As at ambient pressure, it first transforms to the c-As phase at $25\pm1$ GPa \cite{Beister, SilasP}, then to an incommensurate host-guest arsenic (hg-As) phase at a higher pressure of $48\pm11$ GPa \cite{GreeneRG, groupV}. However, Kikegawa and Iwasaki showed that the first transition from g-As to c-As phase occurs in the interval of $31.4-36.6$ GPa, and the c-As remains stable at least up to 45 GPa \cite{KikegawaT}. While the exact pressure from g-As to c-As phase is still debatable, the trend of $T${$\rm_c$} versus pressure is generally consistent across the studies \cite{ChenAL, KawamuraH}. Kawamura and Wittig simply mentioned that the $T${$\rm_c$} of g-As increases monotonously from below 0.05 K at 10 GPa to a maximum of 2.7 K at $\sim$24 GPa \cite{KawamuraH}. Chen $et$ $al.$ reported that the $T${$\rm_c$} of g-As shows a peak value of 2.5 K at 32 GPa associated with the g-As to c-As structural transition \cite{ChenAL}.

\begin{figure}
	\includegraphics[width=8.3cm]{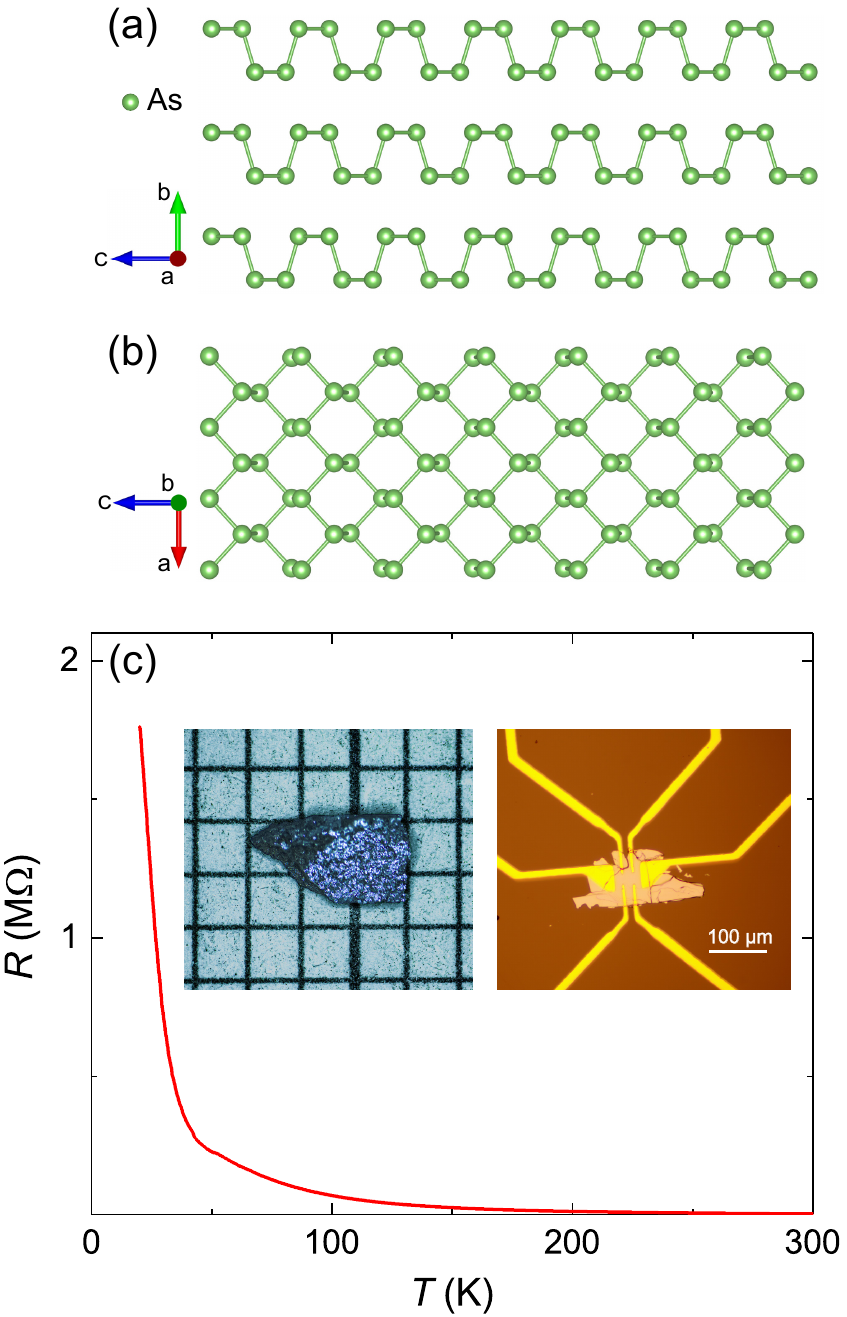}
	\caption{(a) Side view of the atomic structure of b-As. The puckered layer is stacked along the $b$ axis. (b) Top view of the atomic structure of monolayer b-As. (c) Temperature dependence of resistance of a 139-nm b-As thin flake on a sapphire substrate. The $R(T)$ curve shows a typical semiconducting behavior. Left inset: Optical image of a b-As single crystal on the 1 $\times$ 1 mm$^2$ grid. Right inset: Optical image of a typical multi-ternimal b-As flake device, used for resistance measurement under ambient pressure. The thickness of the flake is 139 nm.}
\end{figure}


In this paper, we report high-pressure Raman spectra and resistance measurements on b-As, as well as calculations of phonon density of states (DOS) and pressure-dependent enthalpy of four reported arsenic phases. The evolution of Raman spectra reveals a sequence of pressure-induced structural transitions: from b-As to g-As phase at 1.51 GPa, then to c-As phase at 25.9 GPa, and finally to incommensurate hg-As phase at 44.8 GPa. These structural transitions align with the lowest-enthalpy transitions predicted by theoretical calculations. Remarkably, the c-As and incommensurate hg-As phases manifest superconductivity under pressure. After the emergence of superconducting transition at 25.4 GPa, the $T${$\rm_c$} for the entire c-As phase is around 3 K. Upon further compression, we observe a higher-$T${$\rm_c$} plateau around 4.5 K above 43 GPa in the incommensurate hg-As phase. With these findings, we construct the latest structural and superconducting phase diagrams of b-As under pressure.

\section{Experimental details}

The b-As single crystals used in this work were from natural mineral. It can be mechanically exfoliated into thin flakes. The thickness of the b-As flake on sapphire substrate was measured by a stylus profilometer (Bruker DektakXT). The pressure was generated by a diamond anvil cell (DAC) which has two opposite diamond anvils with a culet diameter of 300 $\mu$m. A stainless-steel DAC was used for Raman spectra measurements. Stainless-steel plate was used as gasket and silicone oil served as an inert pressure transmitting medium (PTM). The selected b-As flake was transferred to the diamond culet by PDMS dry transfer method. Room-temperature Raman spectra were detected by a confocal Raman microscope (Horiba Jobin-Yvon Labram HR Evolution) and the 532 nm laser was available for excitation. A Cu-Be alloy DAC was used for resistance measurements. Cu-Be plate was used as gasket and a mixture of epoxy and powdered cubic boron nitride served as insulating material. Small fragment of shiny b-As single crystal was loaded into the DAC and no PTM was used. The standard van der Pauw four-probe method was adopted to measure the resistance and the four electrodes were made of 4-$\mu$m-thick platinum foil. Low-temperature resistance measurements were carried out in a physical property measurement system (PPMS, Quantum Design). The pressure was measured using a ruby fluorescence calibrant at room temperature for all experiments.

Our calculations are performed using density functional theory as implemented in the Vienna $ab$-$initio$ simulation package (VASP) \cite{Theo1}. We choose the projector-augmented-wave (PAW) method for the basis set \cite{Theo2}. The exchange-correlation functional is treated by the Perdew-Burke-Ernzerhof (PBE) \cite{Theo3} form of the generalized gradient approximation (GGA). The energy cutoff of the plane waves is set to 550 eV. The Brillouin zone (BZ) is integrated using a Gamma-centered Monkhorst-Pack k-mesh method \cite{Theo4} and keep the fine k-mesh spacing smaller than 0.02 \AA$^{-1}$. The convergence criterion is set to 10$^{-7}$ eV/cell in energy and 10$^{-3}$ eV/{\AA} in forces. The phonon DOS are simulated by using the finite displacement method \cite{Theo5} as implemented in the Phonopy programme \cite{Theo6}. The $ab$-$initio$ molecular dynamics (AIMD) simulations of 108 As atoms is modeled with the NPT ensemble under the pressure of 30 GPa. The temperature is set to fluctuation at around 300 K using the Langevin thermotat \cite{Theo7}. The simulation duration is set to 6 picoseconds with a time step of 2 femtoseconds.

\section{Results and discussion}

Layered b-As possesses an orthorhombic crystal structure with a space group of $Cmce$ \cite{ZhongM2}, closely resembling the structure of black phosphorus. As depicted in Figs. 1(a) and 1(b), each arsenic atom within a single layer is covalently bonded with three adjacent arsenic atoms to form a puckered honeycomb arrangement. The puckered layers in b-As are stacked along the $b$ axis by weak van der Waals interactions. The layered crystal structure allows for mechanical cleavage, facilitating the preparation of b-As flakes with different layer numbers \cite{ZhongM, ChenY, ZhongM2, WangY, ShengF}. The left inset of Fig. 1(c) presents the image of a b-As single crystal, while a mechanically cleaved 139-nm b-As flake on a sapphire substrate is shown in the right inset. Figure 1(c) also plots the temperature dependence of the resistance $R(T)$ curve for the b-As flake under ambient pressure, which shows a typical semiconducting behavior.

The Raman spectra of a $\sim$150-nm-thick b-As flake upon compression from 0 to 58.1 GPa at room temperature are presented in Fig. 2(a). At 0 GPa, b-As exhibits three prominent Raman characteristic peaks at 221.2, 226.8, and 254.5 cm$^{-1}$, which correspond to $A_g^1$, $B_{2g}$, and $A_g^2$ modes, respectively. With increasing the pressure to 0.38 GPa, the $A_g^1$ and $B_{2g}$ peaks merge together, resulting in the observation of only two peaks: $A_g^2$ and the merged $A_g^1/B_{2g}$ modes. At 1.51 GPa, besides the peaks of b-As, two new Raman characteristic peaks $E_g$ and $A_{1g}$ for g-As phase are identified at 186.7 and 250.3 cm$^{-1}$, respectively, indicating the initial structural phase transition from b-As to g-As \cite{LiRP, Gao2020}. The b-As and g-As phases coexist in the pressure range from 1.51 to 4.09 GPa. At 4.92 GPa, the $A_g^1/B_{2g}$ and $A_g^2$ modes for b-As vanish, whereas the $E_g$ and $A_{1g}$ modes for g-As persist, showing that b-As completely transforms into g-As phase. As the pressure increases from 1.51 to 24.3 GPa, both the $E_g$ and $A_{1g}$ modes exhibit red shifts with noticeable broadening of peaks. These variations in the Raman peaks for b-As under pressures are consistent with previous report \cite{Gao2020}. Subsequently, the $E_g$ and $A_{1g}$ modes for the g-As phase vanish above 25.9 GPa and no significant peak is detected, suggesting the second structural transition to c-As phase \cite{Du2023}, although previous high-pressure study shows a very weak mode in the c-As phase \cite{Beister}. At a higher pressure of 44.8 GPa, two new modes, NM1 and NM2, appear at 216.6 and 263.2 cm$^{-1}$, respectively, representing the third structural transition to incommensurate hg-As phase \cite{groupV}. Upon further compression up to 58.1 GPa, both NM1 and NM2 exhibit blue shifts.

\begin{figure}
	\includegraphics[clip,width=8.8cm]{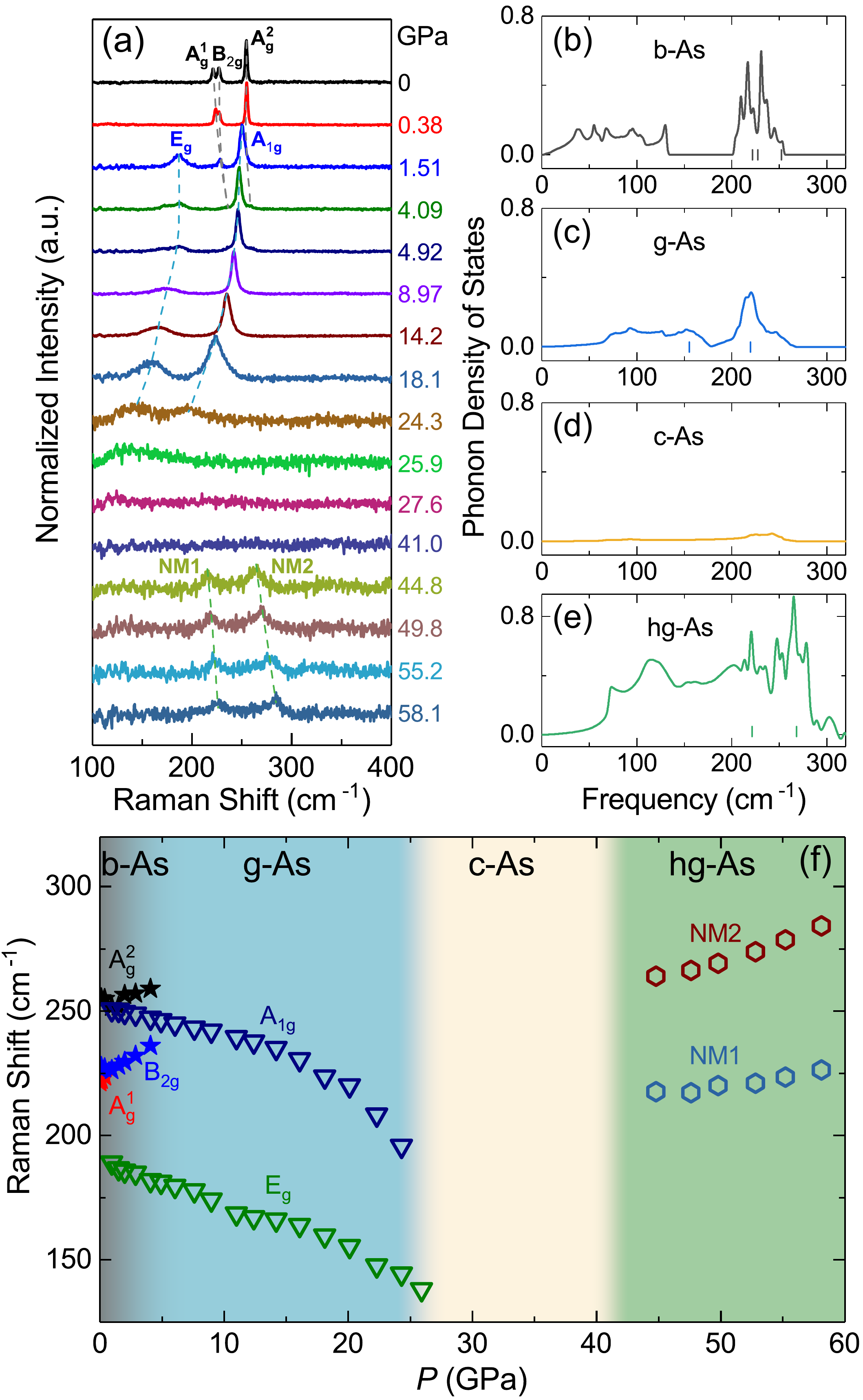}
	\caption{(a) Pressure-dependent Raman spectra of a $\sim$150-nm-thick b-As flake upon compression from 0 to 58.1 GPa at room temperature. Calculated phonon DOS of (b) b-As, (c) g-As, (d) c-As, and (e) incommensurate hg-As phases are shown. The vertical symbols in (b), (c), and (e) represent the experimentally observed Raman shifts for b-As (0 GPa), g-As (20.1 GPa), and incommensurate hg-As (49.8 GPa) phases, respectively. (f) Detected Raman shifts as a function of pressure. At 0 GPa, three Raman peaks $A_g^1$, $B_{2g}$, and $A_g^2$ for b-As are observed. At 1.51 GPa, two new Raman peaks $E_g$ and $A_{1g}$ for g-As phase show up. No significant signals are detected above 25.9 GPa, suggesting the transition to c-As phase. At 44.8 GPa and above, two new modes (NM1 and NM2) for incommensurate hg-As phase appear.}
\end{figure}


\begin{figure}
	\includegraphics[clip,width=8.6cm]{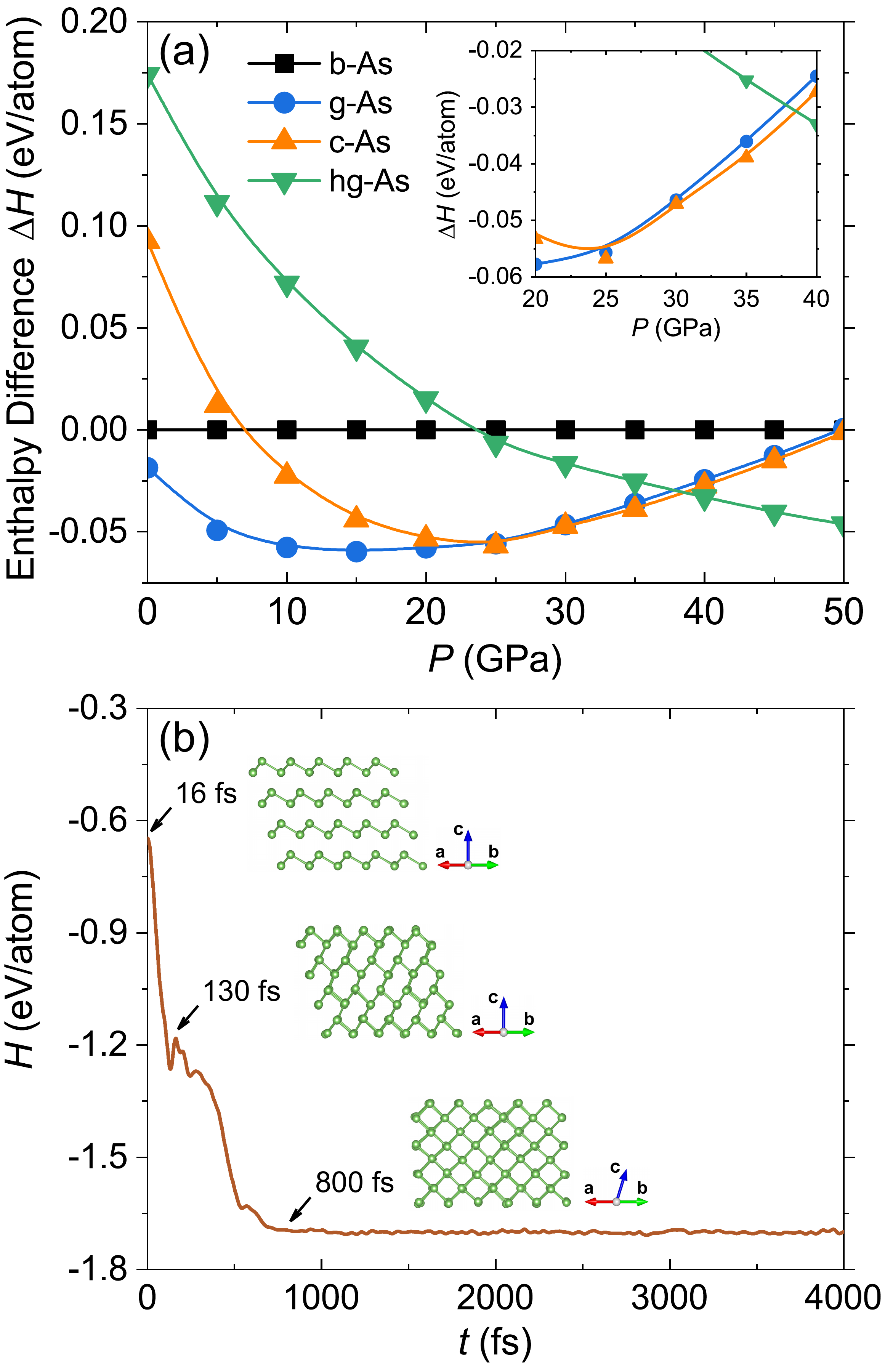}
	\caption{(a) Calculated enthalpy differences per atom of g-As, c-As and incommensurate hg-As phases with respect to b-As as a function of pressure. Inset: Enlargement of the enthalpy differences in the pressure range of $20-40$ GPa. (b) Molecular dynamics simulations for the enthalpy per atom of g-As phase over time at 30 GPa and 300 K, which reveal a spontaneous structural transition from g-As to c-As phase. Insets from the top down depict the crystal structures at 16, 130, and 800 fs. At 16 fs, the crystal structure of g-As phase is still visible. At 130 fs, half of the intralayer As-As bonds has formed, and at 800 fs, the structure has fully transformed to the c-As phase.}
\end{figure}


To better understand the structural phase transitions suggested by Raman spectroscopy, we calculate phonon DOS of different arsenic structural phases under pressure, as shown in Figs. 2(b)-2(e). In Fig. 2(b), at 0 GPa, b-As exhibits three significant peaks in its phonon DOS within the range of 200 to 300 cm$^{-1}$, corresponding to the three vibrational mode peaks observed experimentally. In Fig. 2(c), at 15 GPa, some peaks disappear in the phonon DOS of g-As, along with red shift of the $E_g$ and $A_{1g}$ peaks upon compression. In Fig. 2(d), at 35 GPa, c-As has no prominent peaks in its phonon DOS. Lastly, in Fig. 2(e), at 50 GPa, hg-As displays two peaks near 220 and 260 cm$^{-1}$ in its phonon DOS, which correspond to the experimentally observed NM1 and NM2 peaks.

The pressure dependence of the Raman shift is summarized in Fig. 2(f). For b-As under pressure, both the $A_g^1/B_{2g}$ and $A_g^2$ modes have slight red shifts, followed by blue shifts between $0.38-4.09$ GPa. The rates of blue shift with pressure for the $A_g^1/B_{2g}$ and $A_g^2$ modes are 3.06 and 2.44 cm$^{-1}$/GPa, respectively. For the g-As phase, the Raman characteristic frequencies for both $E_g$ and $A_{1g}$ modes decrease significantly with pressure. The red-shift rate for the $E_g$ mode is $-$1.77 cm$^{-1}$/GPa from 1.51 to 14.2 GPa and $-$2.61 cm$^{-1}$/GPa from 18.1 to 24.3 GPa. The $A_{1g}$ mode first decreases with a rate of $-$1.16 cm$^{-1}$/GPa below 14.2 GPa, and then decreases more rapidly to $-$5.78 cm$^{-1}$/GPa between 18.1 and 24.3 GPa. No significant signals are found in the c-As phase. For the incommensurate hg-As phase, both NM1 and NM2 show blue shifts, with frequencies increasing by 0.68 and 1.55 cm$^{-1}$/GPa, respectively.

\begin{figure*}
	\includegraphics[clip,width=17.8 cm]{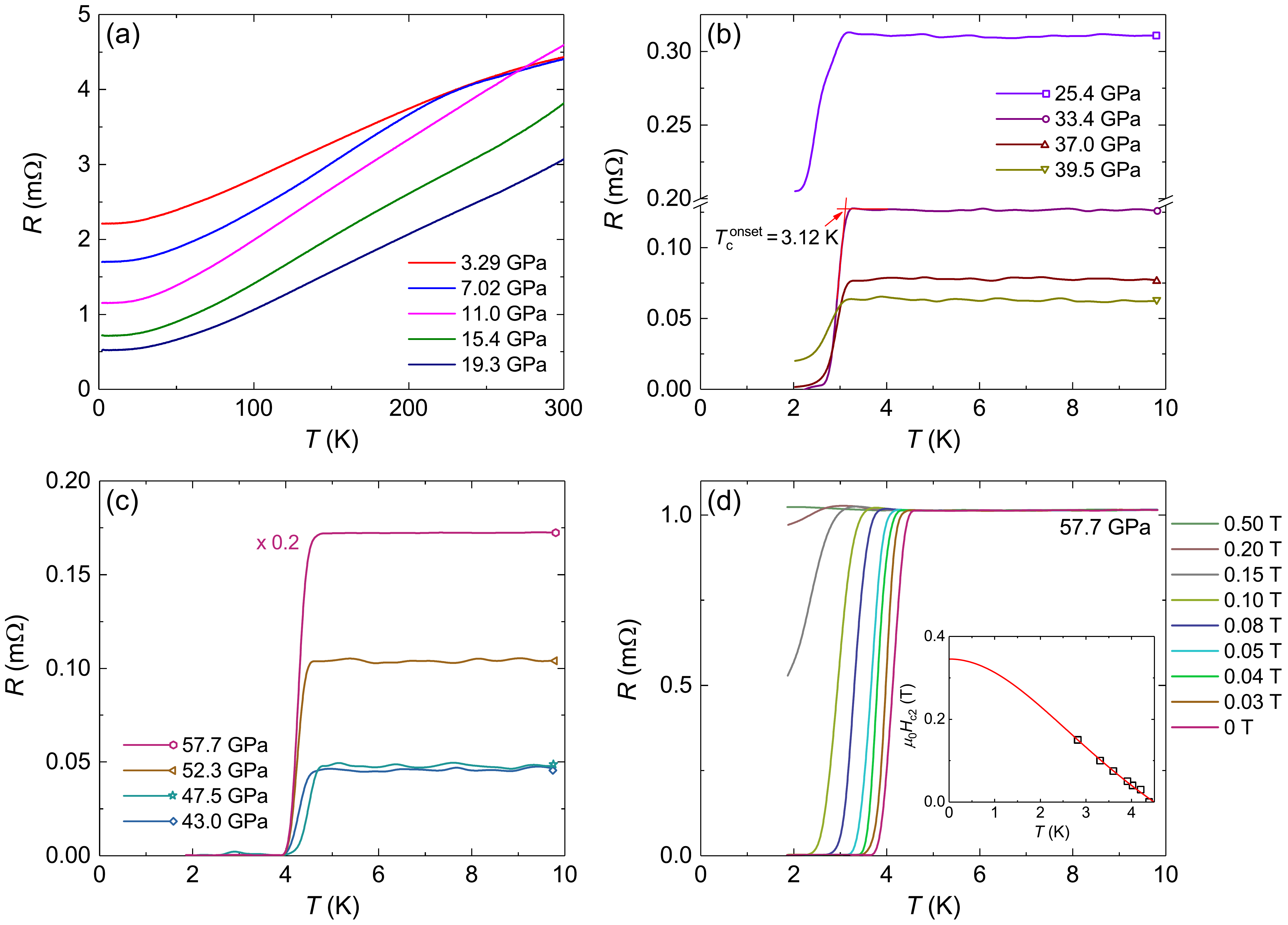}
	\caption{(a) Temperature dependence of resistance for b-As under pressures from 3.29 to 19.3 GPa. Low-temperature superconducting transitions under pressures between (b) $25.4-39.5$ GPa and (c) $43.0-57.7$ GPa. The $T${$\rm_c$} is defined at the onset of resistance drop ($T${$\rm_c$}{$^{\rm{onset}}$}), as indicated on the curve of 33.4 GPa. (d) Low-temperature superconducting transitions in different magnetic fields under the pressure of 57.7 GPa. Inset: Temperature dependence of the upper critical field $\mu_0H{\rm_{c2}}$ under 57.7 GPa. The red line represents the fit to the Ginzburg-Landau formula, $\mu_0H{\rm_{c2}}(T)=\mu_0H{\rm_{c2}}(0)(1-(T/T{\rm_c})^2)/(1+(T/T{\rm_c})^2)$.}
\end{figure*}


Figure 3(a) shows the calculated enthalpy differences per atom for other three structural phases of arsenic referenced to the b-As as a function of pressure. According to the calculations, the enthalpy of the g-As phase is lower than that of the b-As phase between $0-25$ GPa. This responds to the fact that b-As is a metastable phase of arsenic under ambient pressure, while g-As is more favorable, and that the structural transition from b-As to g-As can occur under low pressure. At pressures above 25 GPa, the enthalpy of the c-As phase is the lowest and comparable to that of the g-As phase. The AIMD simulations provide insight into the structural phase transition from g-As to c-As at 30 GPa and 300 K. As shown in Fig. 3(b), the process of this structural transition clearly shows two steps: the nearest-neighbor atoms between layers bond first, and then the rest of the atoms bond, forming a cubic-like structure. This transition occurs spontaneously, with the g-As eventually relaxing to the c-As structural phase under pressure, resulting in the comparable enthalpies for both and the enthalpy of c-As being slightly lower than that of g-As. At pressures between 38.5 and 50 GPa, the enthalpy of the incommensurate hg-As phase becomes lowest, signifying the transition to the hg-As phase. This sequence of lowest-enthalpy transitions matches the structural phase transitions observed in our Raman experiments, as shown in Fig. 2(a).

Figure 4(a) presents the temperature dependence of resistance for b-As under pressures from 3.29 to 19.3 GPa. Within this pressure range, the sample becomes more and more metallic. In Fig. 4(b), a superconducting transition with a resistance drop emerges at 25.4 GPa, near the critical pressure of g-As to c-As structural transition, and the superconducting transition with zero resistance is observed upon further compression. The low-temperature superconducting transitions under high pressures between $25.4-39.5$ GPa and $43.0-57.7$ GPa are plotted in Figs. 4(b) and 4(c), respectively. The $T${$\rm_c$} is defined at the onset of the resistance drop ($T{\rm_c}^{\rm{onset}}$), as indicated on the $R$($T$) curve of 33.4 GPa. With increasing the pressure, $T${$\rm_c$} slightly rises from 3.06 K at 25.4 GPa to 3.12 K at 33.4 GPa, then decreases from 3.12 K at 37.0 GPa to 3.04 K at 39.5 GPa, across the whole c-As phase. Here, the $T${$\rm_c$} versus pressure trend is similar to the previously reported results \cite{KawamuraH, ChenAL}. Interestingly, as the pressure increases to 43.0 GPa, a higher $T_{\rm c}$ of 4.49 K is present, then further goes to 4.70 K at 47.5 GPa, and finally stabilizes at 4.50 K up to the highest measured pressure of 57.7 GPa. Such a higher $T${$\rm_c$} has not been reported in Refs. \cite{KawamuraH, ChenAL}, which is very probably due to the formation of the incommensurate hg-As phase under high pressures \cite{GreeneRG, Tsuppayakorn}, as evidenced by the Raman spectra (Fig. 2(a)). As shown in Fig. 4(c), the resistance of the incommensurate hg-As phase starts to increase with pressure. However, the superconducting transition with nonzero resistance is likely caused by some degree of pressure inhomogeneity \cite{ChiZ2018}.

\begin{figure}
	\includegraphics[clip,width=8.5cm]{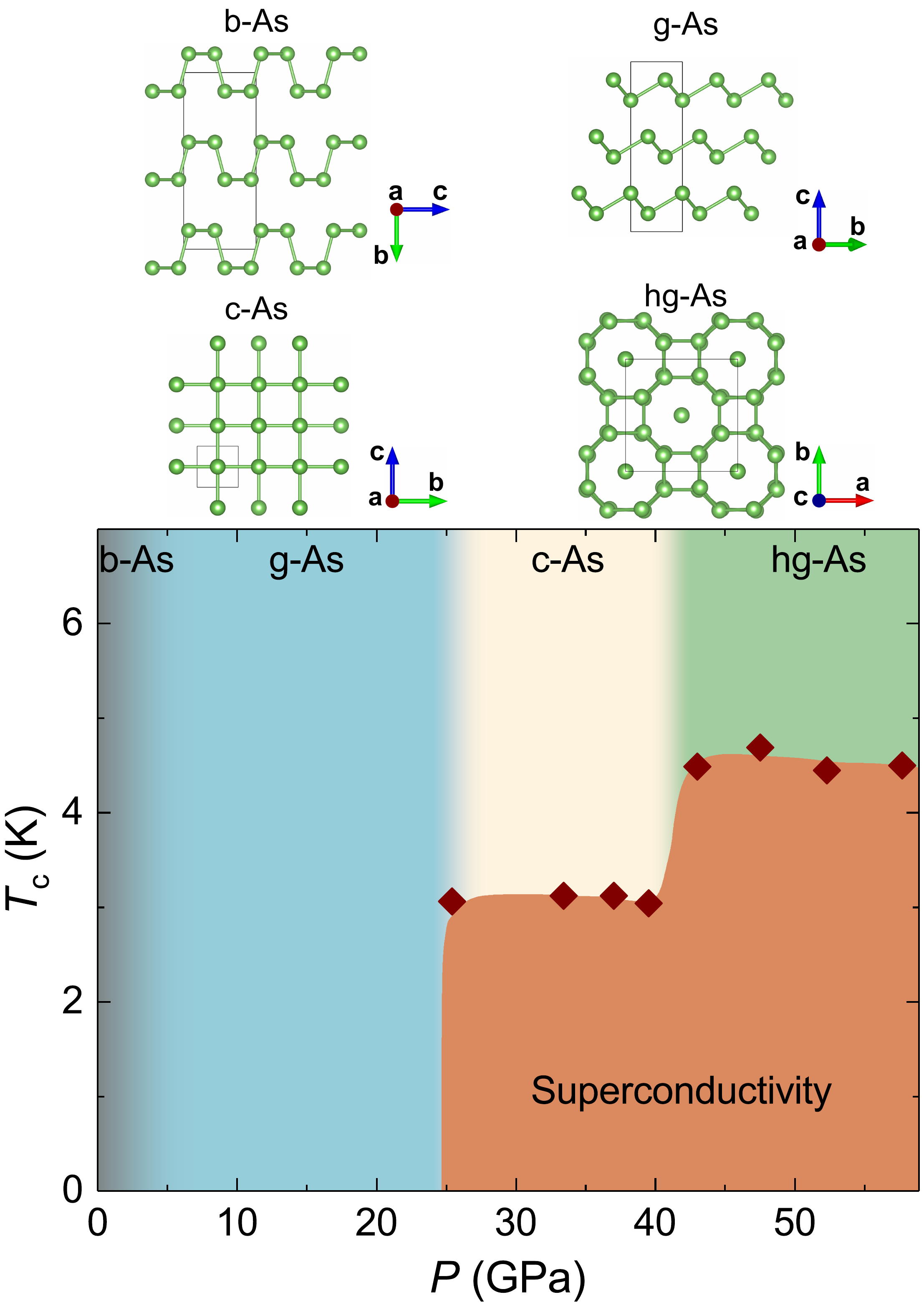}
	\caption{Temperature-pressure superconducting phase diagram of b-As. The crystal structures of b-As, g-As, c-As, and incommensurate hg-As phases are shown in the upper part of the panel. There is a superconducting plateau near 3 K at 25.4 GPa and above, across the entire c-As phase. A higher-$T${$\rm_c$} superconducting plateau around 4.5 K is observed above 43 GPa in the incommensurate hg-As phase.}
\end{figure}

As shown in Fig. 4(d), the pressure-induced superconductivity in b-As is confirmed by the resistance measurements in different magnetic fields under the pressure of 57.7 GPa. As expected, the superconducting transition is suppressed to lower temperatures with increasing field and completely disappears in 0.50 T. The temperature dependence of the upper critical field $\mu_0H{\rm_{c2}}$ is plotted in the inset of Fig. 4(d). The Ginzburg-Landau equation, $\mu_0H{\rm_{c2}}(T)=\mu_0H{\rm_{c2}}(0)(1-(T/T{\rm_c})^2)/(1+(T/T{\rm_c})^2)$, is used to estimate the $\mu_0H{\rm_{c2}}(0)$, which yields $\mu_0H{\rm_{c2}}(0) \approx$ 0.35 T.

Based on above high-pressure Raman and resistance measurements of b-As, as well as pressure-dependent enthalpy difference calculations for different arsenic phases, we construct the structural and superconducting phase diagrams of b-As under pressure, shown in Fig. 5. Superconductivity emerges at 25.4 GPa, with $T${$\rm_c$} of 3.1 K, and exhibits a superconducting plateau across the entire c-As phase. The $T${$\rm_c$} near 3 K is similar to earlier report \cite{KawamuraH, ChenAL}. Beyond this superconducting plateau, a higher-$T${$\rm_c$} plateau emerges at 4.5 K starting at 43 GPa, associated with the c-As to incommensurate hg-As structural phase transition \cite{GreeneRG,Tsuppayakorn}. Previous experimental studies have shown that the maximum $T${$\rm_c$} for g-As under pressure is 2.7 K at approximately 24 GPa \cite{ChenAL, KawamuraH, Beister}, while the theoretical calculation has predicted a $T${$\rm_c$} of 4.2 K in a body-centered tetragonal structure of arsenic at 150 GPa \cite{Tsuppayakorn}. Our current work has updated the superconducting phase diagram of arsenic under pressure, starting from b-As. Although our study does not include susceptibility measurements to confirm the Meissner effect in b-As under pressure, this limitation represents an area for future efforts to definitively claim the superconductivity in this compound.

\section{Conclusion}
We have experimentally investigated the physical properties of the novel semiconductor b-As by measuring Raman spectra and resistance under pressure up to 58 GPa. According to the Raman spectra, a structural phase transition from b-As to g-As occurs at a pressure of 1.51 GPa, with both phases coexisting in a pressure range of $1.51-4.09$ GPa. The b-As completely transforms into the g-As phase at 4.92 GPa and then to the c-As and the incommensurate hg-As phases at 25.9 and 44.8 GPa, respectively. These critical pressures for the phase transitions are consistent with the results of the lowest-enthalpy transitions predicted by theoretical calculations. Superconductivity with $T${$\rm_c$} of 3.1 K is induced at 25.4 GPa. From 25.4 to 39.5 GPa in the c-As phase, $T${$\rm_c$} shows a superconducting plateau around 3 K. Beyond this plateau, further compression induces a steep increase in $T${$\rm_c$} to 4.5 K at 43 GPa, after which $T${$\rm_c$} remains nearly constant in the incommensurate hg-As phase. Our study provides a better understanding of the pressure-induced structural phase transitions and the associated superconductivity in the novel semiconductor, b-As. We present a significantly updated phase diagram of the superconducting state in b-As under pressure up to 58 GPa.

~\\

\begin{acknowledgments}
We thank Jing Wang and Zeyuan Sun for helpful discussions. Part of the experimental work has been carried out in Fudan Nanofabrication Laboratory. This work was supported by the Natural Science Foundation of China (Grant No. 12174064), the Shanghai Municipal Science and Technology Major Project (Grant No. 2019SHZDZX01), the Innovation Program for Quantum Science and Technology
(Grant No. 2024ZD0300104), the National Key Research and Development Program of China (Grant Nos. 2021YFA1400100 and 2021YFA1400500), and the Strategic Priority Research Program of the Chinese Academy of Sciences (Grant No. XDB33000000).
\end{acknowledgments}

~\\
\noindent $^*$ These authors contributed equally to this work.\\
\noindent $^\dag$ E-mail: syhuang@fudan.edu.cn\\
\noindent $^\ddag$ E-mail: shiyan$\_$li$@$fudan.edu.cn

\end{document}